\documentclass[iop,apj,tighten]{emulateapj}
\usepackage[breaklinks,colorlinks,urlcolor=blue,citecolor=blue,linkcolor=blue]{hyperref}
\usepackage{amsmath,amssymb}
\usepackage{cleveref}
\usepackage{graphicx}
\usepackage{bm}

\usepackage{appendix}
\usepackage{color}
\newcommand{\Msun}{{\rm M}_\odot}

\shorttitle{Mass-gap mergers in AGN disks}
\shortauthors{Tagawa et al.}

\begin{document}
\title{Mass-gap Mergers in Active Galactic Nuclei}

\author{Hiromichi Tagawa\altaffilmark{1}, Bence Kocsis\altaffilmark{2}, Zolt{\'a}n Haiman\altaffilmark{3}, Imre Bartos\altaffilmark{4}, 
Kazuyuki Omukai\altaffilmark{1},
Johan Samsing\altaffilmark{5}
}
\affil{\altaffilmark{1}Astronomical Institute, Graduate School of Science, Tohoku University, Aoba, Sendai 980-8578, Japan\\
\altaffilmark{2} Rudolf Peierls Centre for Theoretical Physics, Clarendon Laboratory, Parks Road, Oxford OX1 3PU, UK\\
\altaffilmark{3}Department of Astronomy, Columbia University, 550 W. 120th St., New York, NY, 10027, USA\\
\altaffilmark{4}{Department of Physics, University of Florida, PO Box 118440, Gainesville, FL 32611, USA}\\
\altaffilmark{5}Niels Bohr International Academy, The Niels Bohr Institute, Blegdamsvej 17, 2100 Copenhagen, Denmark
}
\email{E-mail: htagawa@astr.tohoku.ac.jp}

\begin{abstract} 
The recently discovered gravitational wave sources GW190521 and GW190814 have shown evidence of BH mergers with masses and spins outside of the range expected from isolated stellar evolution. These merging objects could have undergone previous mergers. Such hierarchical mergers are predicted to be frequent in active galactic nuclei (AGN) disks, where binaries form and evolve efficiently by dynamical interactions and gaseous dissipation. Here we compare the properties of these observed events to the theoretical models of mergers in AGN disks, which are obtained  by performing one-dimensional $N$-body simulations combined with semi-analytical prescriptions. 
The high BH masses in GW190521 are consistent with mergers of high-generation (high-g) BHs where the initial progenitor stars had high metallicity, 2g BHs if the original progenitors were metal-poor, or 1g BHs that had gained mass via super-Eddington accretion. Other measured properties related to spin parameters in GW190521 are also consistent with mergers in AGN disks. 
Furthermore, mergers in the lower mass gap or those with low mass ratio as found in GW190814 and GW190412 are also reproduced by mergers of 2g-1g or 1g-1g objects with significant accretion in AGN disks. Finally, due to gas accretion, the massive neutron star merger reported in GW190425 can be produced in an AGN disk. 
\end{abstract}
\keywords{
binaries: close
-- gravitational waves 
--galaxies: active
-- methods: numerical 
-- stars: black holes 
}

\section{Introduction}

Recently, several gravitational wave (GW) events were reported by 
LIGO and Virgo whose measured physical properties pose interesting constraints on their astrophysical origin.  
One of these is GW190521 \citep{LIGO20_GW190521,LIGO20_GW190521_astro}, in which the masses of the merging BHs ($85^{+21}_{-14}\,\Msun$ 
and $66^{+17}_{-18}\,\Msun$, but see \citealt{Nitz20}) indicate that the merging BHs fall into the so-called ``upper-mass gap'', as they are heavier than the maximum BH mass imposed by (pulsation) pair-instability supernovae \citep{Chatzopoulos12,Farmer19}. The components' dimensionless spins ($a=0.69^{+0.27}_{-0.62}$ and $0.73^{+0.24}_{-0.64}$) are also larger than the maximum natal spins predicted by stellar evolutionary models for angular momentum transfer \citep{Fuller19_massive}. 
A possible electromagnetic counterpart has been reported for GW190521, which, if real, would suggest mergers in gas-rich environments \citep{2019ApJ...884L..50M,Graham20}. 
Further candidates for mergers in the upper-mass gap are reported in GW170729 \citep{TheLIGO18}, GW170817A \citep{Zackay19_GW170817A}, and six other candidates in O3 \citep{LIGO20_O3_Catalog}. 
A possible explanation proposed for the large masses in these events is isolated binary evolution, presuming that massive BHs can form from Population~III (Pop~III) or metal-poor stars \citep{Tanikawa20_GW190521,Kinugawa20_GW190521,Farrell20,Liu20_GW190521_pop3,Liu20_GW190521_dynamical,Belczynski20_GW190521}.
However, due to their expected rapid rotation, stellar remnants are unlikely to retain their hydrogen envelopes and produce BHs as massive as in GW190521 \citep{Chatzopoulos12,Yoon12}. Likewise, 
the probability that the progenitor of the $85\,\Msun$ BH avoided a pulsation pair-instability supernova due to uncertainties on the reaction rate of carbon burning is low \citep{Farmer19}. 
An alternative proposal is for Pop~III stellar remnants to grow via accretion~\citep{Safarzadeh20_GW190521}; this requires the remnant BHs to spend a significant time in a dense gaseous environment of their host protogalaxies. 

Similarly, the neutron star (NS) merger GW190425 indicates masses higher than in NS binaries observed in the Milky Way \citep{LIGO20_GW190425}. 
To explain this event, a rapid merging pathway is proposed for the binary evolution channel \citep{Romero-Shaw20_GW190425,Safarzadeh20_GW190425,Galaudage20_GW190425}. 
Additionally, mergers with very unequal masses have been reported -- GW190412 \citep{LIGO20_GW190412} and GW190814 \citep{LIGO20_GW190814} --, which are also atypical in stellar evolutionary models of isolated binaries 
\citep{Gerosa20,Olejak20_GW190412}. 
The mass of the secondary object of $2.6\,\Msun$ in GW190814 falls in the so-called lower mass gap, predicted by stellar-evolutionary models 
(\citealt{Fryer12,Zevin20_GW190814} but see \citealt{Belczynski12,Drozda20}) 
and obeyed by observations of X-ray binaries \citep{Ozel12}. 
The object in the lower mass gap in GW190814 and a non-zero spin for the primary BH ($0.43^{+0.16}_{-0.26}$) in GW190412 are consistent with a scenario in which the merging compact objects (COs) had experienced previous mergers or significant accretion. 
These events suggest that growth by gas accretion or hierarchical mergers may be common among COs \citep[see e.g.][]{OLeary16,Gerosa20,LIGO20_GW190521_astro,Hamers20_GW190412,Rodriguez20_GW190412,Safarzadeh20_GW190814,Safarzadeh20_GW190521,Yang20_GW190814,LiuLai20_GWevents}.

By analyzing the ensemble of events detected during LIGO/Virgo's O1-O3a observing runs, 
\cite{2020arXiv201105332K} and \cite{Tiwari20} found preference 
for at least one, but probably multiple hierarchical mergers in the detected sample. The results by \cite{2020arXiv201105332K} strongly depend on the escape velocity at the merger sites, with higher escape velocities favoring a larger number of hierarchical mergers. 

Active galactic nuclei (AGNs) represent environments where mass growth by gas accretion is possible~\citep{Levin2007} 
and hierarchical mergers are frequent due to efficient hardening by interaction with gas \citep[e.g.][]{Bartos17,Stone17,McKernan17,McKernan20_Monte,McKernan20_NSWD}, stars, and COs \citep[][hereafter Paper~I]{Tagawa19}. Additionally, the high escape velocity in such systems helps retain recoiling merger remnants. 
\citet{Yang20_gap} investigated the possibility of lower mass gap mergers by assuming that the objects are quickly delivered to migration traps of AGN disks where they undergo mergers. 
In Paper~I, we performed one-dimensional (1D) $N$-body simulations combined with semi-analytical prescriptions for gaseous and stellar effects, to follow the evolution of binary separation, along with the radius and inclination of the orbit around the SMBH. The results showed that binaries form efficiently during dynamical encounters in gas, and 
mergers are facilitated by binary-single (BS) interactions in the gap-forming inner regions. 
In \citet[][hereafter Paper~II]{Tagawa20b_spin}, we extended this model to follow the evolution of BH spins, and binary orbital angular momentum directions. In 
\citet[][hereafter Paper~III]{Tagawa20_ecc}, we also included the effects of binary eccentricity, and found that mergers are often
highly eccentric in AGN disks if BS interactions are confined to the AGN plane due to gaseous torques (see also \citealt{Samsing20}) or if the objects migrate close to the SMBH where they undergo GW capture events. 

In this paper, 
we extend our previous models to include NSs. We then
examine the mass distribution of merging binaries in the AGN channel, and assess the likelihood of mergers in the lower and upper mass gaps, those with highly unequal mass ratios, and massive NS mergers as seen in GW190521, GW190814, GW190412, and GW190425.

\section{Method}

Our model is based on Papers I, II, and III, summarized in \S~\ref{subsec:modelsummary}, with new ingredients described in $\S\,\ref{sec:new_objects}$. 

\subsection{Overview of model}
\label{subsec:modelsummary}

We consider a model with a supermassive BH (SMBH) at the center of a galaxy, surrounded by a gaseous accretion disk (hereafter AGN disk) and a spherical nuclear star cluster (NSC), which are assumed to be in a fixed steady-state.
We follow the $N$-body evolution of COs consisting of 
those in the nuclear stars cluster 
and those captured inside the AGN disk.

We follow several pathways to form binaries.  
Some fraction of COs are initially in binaries. 
In the AGN disk, binaries form due to gas dynamical friction during two-body encounters and dynamical interactions during three-body encounters. 
Highly eccentric CO binaries also form due to the GW-capture mechanism in single-single encounters \citep[e.g.][]{OLeary09} and BS exchange interactions (Paper~III). 

Regarding the interaction with gas, the velocities of all COs relative to the local AGN disk decrease due to accretion torque and gas dynamical friction.
The binaries' semi-major axis decreases due to gas dynamical friction from the AGN disk. Their semi-major axis and eccentricity evolve also due to type I/II migration torque from a circumbinary disk. 
COs migrate toward the SMBH due to type I/II torques from the AGN disk. 
CO masses, BH spins, and the orbital angular momentum directions of binaries 
gradually evolve due gas accretion 
(Paper~II). 

We also account for dynamical interactions with single stars/COs and CO binaries.  The binaries' semi-major axis, velocities, orbital angular momentum directions, and eccentricity evolve due to BS interactions as prescribed in \citet{Leigh18}, and the velocities of all COs additionally evolve due to scattering. 
The soft binaries, i.e. those with binding energy 
lower than that of scattering objects, are either softened or ultimately disrupted during BS interactions. 
Hard binaries become harder during BS interactions, and exchange interactions may take place; the most massive pair is assumed to remain in the binary after the interaction (Paper~III). 

The binaries' separation and eccentricity decrease due to GW emission. 
When the pericenter distance of the binary becomes smaller than the innermost stable orbit, 
we assume that BHs merge, assign a kick velocity and mass loss due to GW emission, and prescribe BH spin evolution at the merger as in Paper~II.

\begin{table*}
\begin{center}
	\caption{
		The results of different models at $t=10\,\mathrm{Myr}$. 
        The two leftmost ("input") columns show the model number and the differences from the fiducial model. 
       The "output" columns list 
       the fraction of mergers weighted by the detectable volume 
       for 
       BH-LGO ($f_\mathrm{BHLGO}$), 
       LGO-LGO ($f_\mathrm{LGOLGO}$), 
       LGO-NS ($f_\mathrm{LGONS}$), 
       BH-BH ($f_\mathrm{BHBH}$), 
       BH-NS ($f_\mathrm{BHNS}$), 
       NS-NS ($f_\mathrm{NSNS}$) binaries, 
       BH-BH binaries with $m_\mathrm{bin}>130\,\Msun$  ($f_\mathrm{M130}$), 
       and the total number of mergers  ($N_\mathrm{mer}$), 
       where LGOs represent lower mass gap objects with masses in $2$--$5\,\Msun$. 
        }
\label{table_results}
\begin{tabular}{c|c||c|c|c|c|c|c|c|c}
\hline
\multicolumn{2}{c}{input} \vline& \multicolumn{8}{c}{output}\\\hline
Model&Parameter
&$f_\mathrm{BHLGO}$
&$f_\mathrm{LGOLGO}$
&$f_\mathrm{LGONS}$
&$f_\mathrm{BHBH}$
&$f_\mathrm{BHNS}$
&$f_\mathrm{NSNS}$
&$f_\mathrm{M130}$
&$N_\mathrm{mer}$
\\\hline

M1&Fiducial&
$4\times 10^{-3}$&$3\times10^{-4}$&$3\times 10^{-4}$
&0.989&$6\times 10^{-3}$&$3\times 10^{-4}$
&$0.03$
&$1.6\times 10^3$
\\\hline

M2&$m_\mathrm{1gBH}~\leq 45\,\Msun$&
$7\times 10^{-4}$&$5\times 10^{-5}$&$7\times 10^{-5}$
&0.998&$1\times 10^{-3}$&$5\times 10^{-5}$
&0.11
&$2.0\times 10^3$
\\\hline

M3&$\Gamma_\mathrm{Edd,cir} =20$&
$5\times 10^{-3}$&$5\times 10^{-5}$&$9\times 10^{-5}$
&0.994&$8\times 10^{-4}$&$5\times 10^{-5}$
&0.35
&$1.9\times 10^3$
\\\hline

\end{tabular}
\end{center}
\end{table*}

\subsection{Evolution of neutron stars}
\label{sec:new_objects}

While only stellar-mass BHs are followed in Papers~I/II/III, 
here we additionally follow the evolution of NSs among the $N$-body particles. 
We assume that stars with zero-age main-sequence masses of $8$--$20\,\Msun$ form NSs with $1.3\,\Msun$. There are initially $5\times 10^4$ NSs in the fiducial model. 
As NSs are lighter than BHs, 
we assume a shallower radial density profile 
\begin{align}\label{eq:bh_density}
\frac{dN_{\rm NS, ini}(r)}{dr} \propto r ^{-0.5}, 
\end{align}
where $N_{\rm NS, ini}(r)$ is the total initial number of NSs within distance $r$ from the central SMBH, 
and a higher velocity dispersion in the frame comoving with the AGN ($\sigma_{\rm NS} = 0.4 v_\mathrm{kep}(r)$, where $v_\mathrm{kep}(r)$ is the Keplerian velocity at $r$) compared to those for BHs ($dN_{\rm BH, ini}(r)/dr \propto r^0$, $\sigma_{\rm BH} = 0.2 v_\mathrm{kep}(r)$),
as expected from relaxation processes \citep{Hopman06,Szolgyen18}. 
For simplicity, we assume that the spin magnitudes of NSs are always zero, the mass lost during mergers is $0.01\,\Msun$, and their radius is $10\,\mathrm{km}$, within which NSs are assumed to merge in addition to the condition described above. 
We regard objects with mass less than $2\,\Msun$ to be NSs, while
we call objects with masses in the range $2$--$5\,\Msun$ \textit{lower mass gap objects} (LGOs), and treat LGOs as BHs in the calculations. 
We assign no kicks during NS-CO mergers for simplicity \citep[but see][]{Dietrich17}, 
while the spin evolution during NS-CO mergers is prescribed as for BH-BH mergers but with zero spin magnitudes for NSs. 
The interactions of NSs with gas, COs, and stars are prescribed 
in the same manner as for BHs.

\begin{figure*}\begin{center}
\includegraphics[width=120mm]{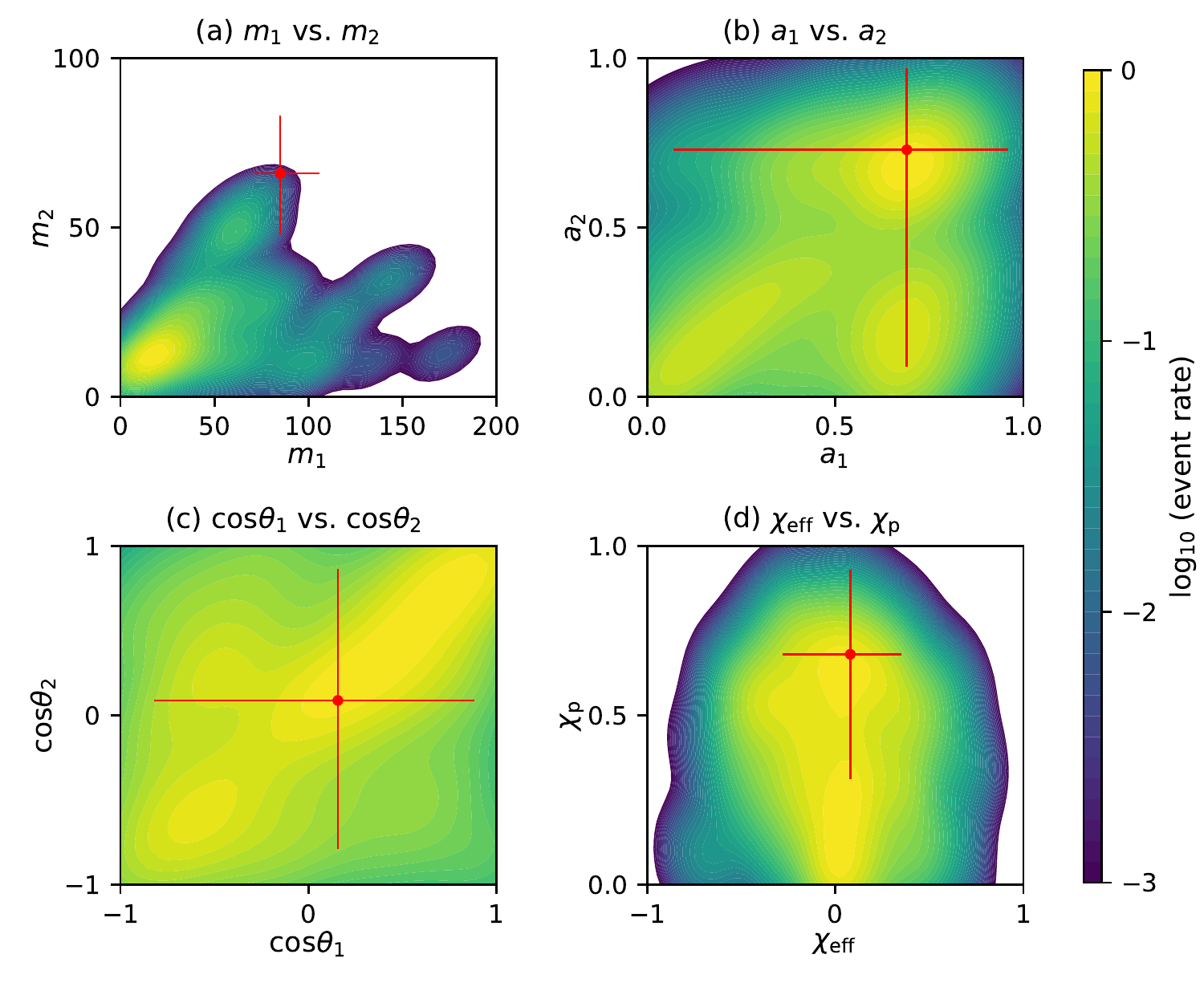}
\caption{
The parameter distributions for merging binaries weighted by the detectable volume at $t=10\,\mathrm{Myr}$ in model~M1. 
The four panels present distributions of (a) the masses of primary and secondary object ($m_1$ {\it vs.} $m_2$) in units of $\Msun$, 
(b) the dimensionless spins of primary and secondary objects ($a_1$ {\it vs.} $a_2$), 
(c) the angles between the binary orbital angular momentum and the BH spins ($\mathrm{cos}\theta_1$ and $\mathrm{cos}\theta_2$), 
and 
(d) the mass-weighted sum of the individual spin components perpendicular to the orbital plane ($\chi_\mathrm{eff}$) {\it vs.} the precession parameter ($\chi_\mathrm{p}$), respectively. 
The error bars correspond to the $90$ percentile credible intervals inferred in GW190521 \citep{LIGO20_GW190521}. 
}
\label{fig:spins_8_fidw}
\end{center}\end{figure*}

\begin{figure*}\begin{center}
\includegraphics[width=180mm]{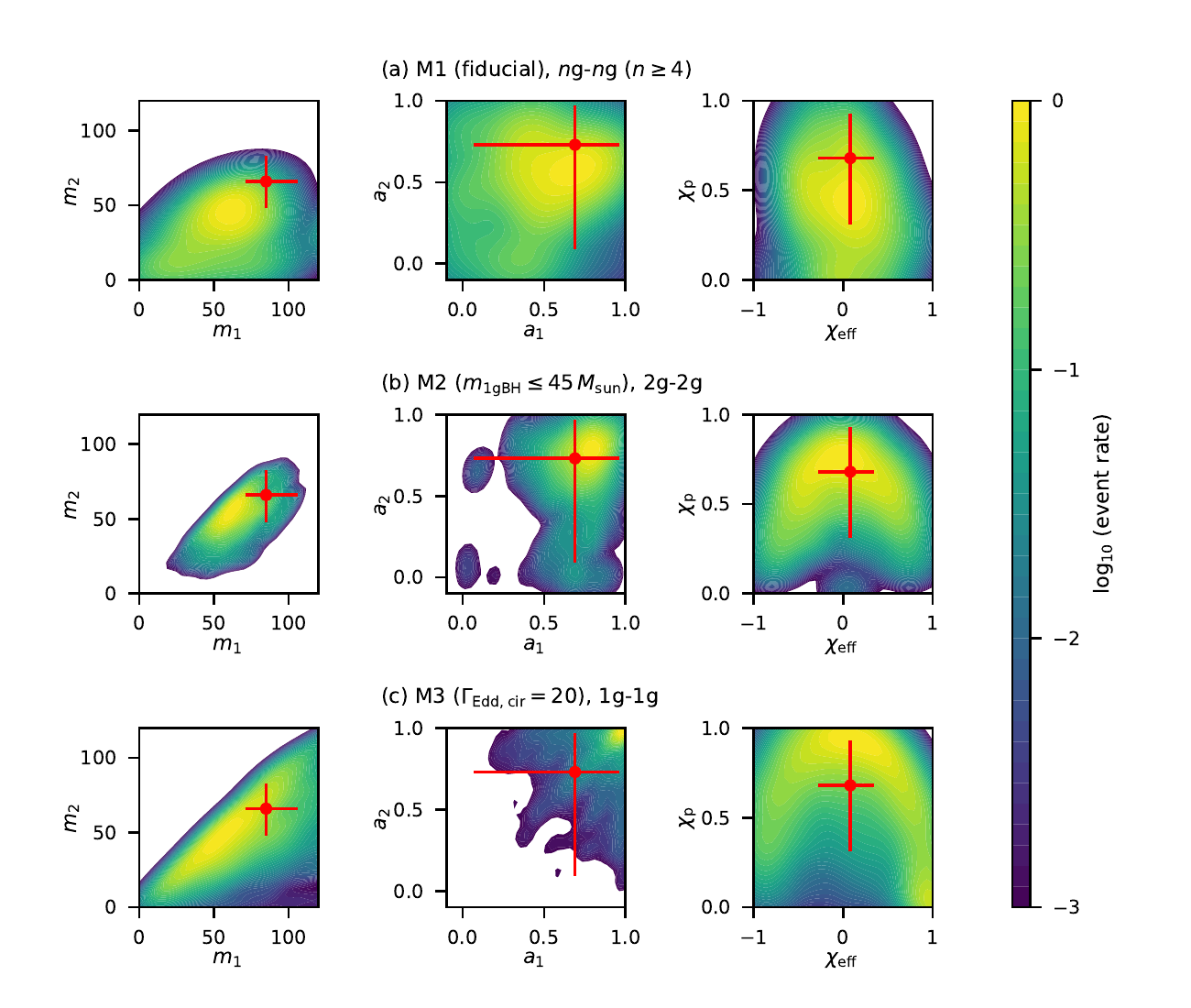}
\caption{
The parameter distributions for merging binaries
weighted by the detectable volume 
among $n$g-$n$g mergers with $n\geq4$ in model~M1 (a), 2g-2g mergers in model~M2 (b), and 1g-1g mergers in model~M3 (c), where $n$g BHs represent BHs 
consisted of merged 
$n$-1g BHs. 
Left, middle, and right panels show the distributions of $m_1$ {\it vs.} $m_2$ in units of $\Msun$, and the dimensionless spin parameters $a_1$ {\it vs.} $a_2$, and $\chi_\mathrm{eff}$ {\it vs.} $\chi_\mathrm{p}$, respectively. 
The error bars correspond to the $90$ percentile credible intervals for GW190521 \citep{LIGO20_GW190521}. 
}\label{fig:masses_gw190521}
\end{center}\end{figure*}

\subsection{Numerical choices}
\label{sec:numerical_choice}

The parameters of the fiducial model (M1) are listed in Table~\ref{table:parameter_model} in the Appendix, which are the same as those in Papers~I/II/III.
We also study the additional models M2 and M3, which differ in the initial masses of stellar-mass BHs and efficiency of gas accretion. 

In model~M1, we set the 
mass of first-generation (1g) BHs 
to $m_\mathrm{1gBH} \leq 15\,\Msun$ given the observed high metallicity population for $90\%$ of stars in the Milky Way NSC \citep[e.g.][]{Do20} 
and the observations indicating strong mass loss during the luminous blue variable phase of solar-metallicity stars \citep[e.g.][]{Chen15}, which likely limits the maximum mass of 1g BHs to be $\sim 15\,\Msun$ \citep[e.g.][]{Belczynski10}. 

On the other hand, in model~M2, the initial masses of 1g BHs are multiplied by 3 compared to the values in model~M1, and so $m_\mathrm{1gBH}\leq45\,\Msun$. 
Such high masses for 1g BHs may exist 
as indicated by a moderately metal-poor subpopulation in the Milky Way's NSC \citep{Do20}, and given that there are significant uncertainties in the initial mass function of 1g BHs in NSCs due to star formation in strong gravitational field \citep[e.g.][]{Nayakshin07}, gas accretion \citep[e.g.][]{Davies20}, 
uncertainties in stellar evolution \citep[e.g.][]{Belczynski10}, and the diversity in the possible origin of objects \citep[e.g.][]{Mapelli16}. 

In model~M3, we increase the maximum accretion rate in Eddington units onto stellar-mass COs to $20$, given the uncertainties due to radiation feedback onto dust \citep{Toyouchi20} 
and rotating gas 
\citep{Jiang+2014,Sugimura18}, and the possible anisotropy of radiation \citep{Proga00}.

\begin{figure*}\begin{center}
\includegraphics[width=180mm]{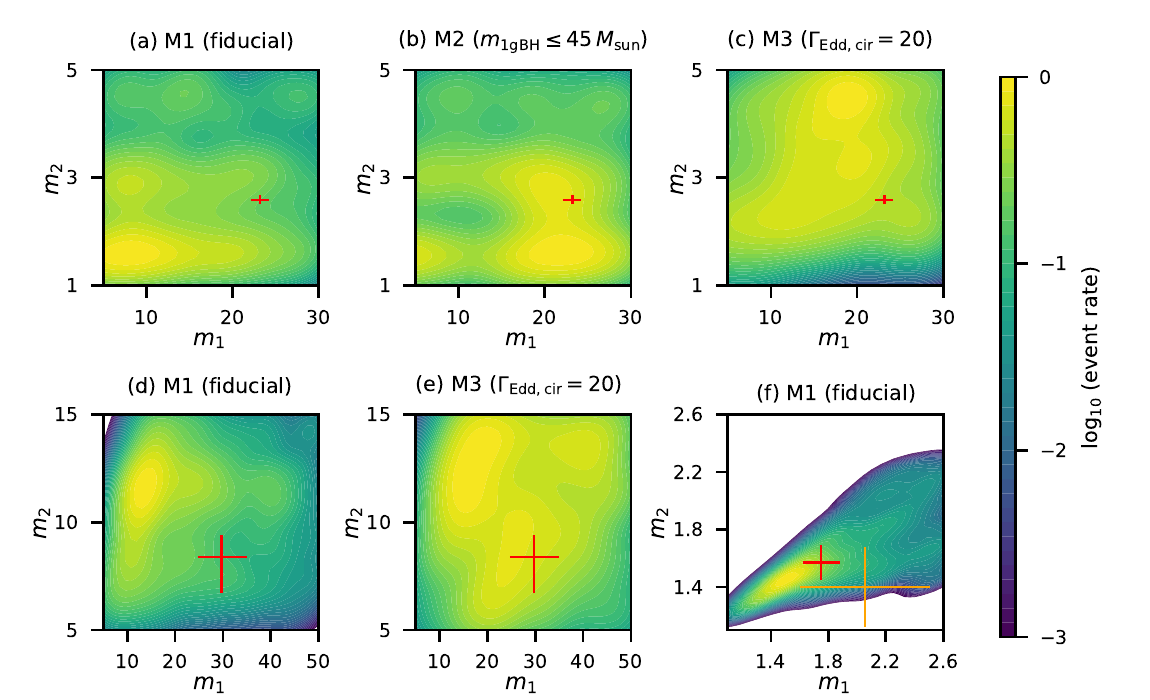}
\caption{
The distribution of $m_1$ and $m_2$ in units of $\Msun$ 
weighted by the detectable volume 
in models~M1--M3 at $t=100\,\mathrm{Myr}$. 
The error bars correspond to the $90$ percentile credible intervals for GW190814 (panels~a--c, \citealt{LIGO20_GW190814}), GW190412 (panels~d and e, \citealt{LIGO20_GW190412}), and GW190425 (panel~f, \citealt{LIGO20_GW190425}) for low-(red) and high-(orange) spin priors. 
}\label{fig:masses_gw190814}
\end{center}\end{figure*}

\section{Results}

\subsection{Upper mass-gap mergers}

We first compare the properties of predicted mergers in AGN disks with those of the upper-mass gap event GW190521. 
The parameter distributions for all events at $t=10\,\mathrm{Myr}$ in model~M1 are shown in Fig.~\ref{fig:spins_8_fidw}, while 
those for specific generation mergers at $t=100\,\mathrm{Myr}$ 
reproducing well GW190521 in models~M1--M3 
in Fig.~\ref{fig:masses_gw190521}. 
Note that the maximum binary masses are not sensitive to time in later phases of $t\geq 10\,\mathrm{Myr}$ as discussed in the Appendix. 
In Fig.~\ref{fig:masses_gw190521}, results at $t=100\,\mathrm{Myr}$ are presented to reduce variation due to a small number of mergers. 
The distributions are weighted by the observable volume 
depending on the masses of merging binaries 
calibrated for aLIGO Handford prior to O3 
as in Paper~I, while the dependence on either $\chi_\mathrm{eff}$ or eccentricity is ignored in terms of detection volume for simplicity. 
The plots are smoothed by performing kernel-density estimate using Gaussian kernels whose bandwidth is chosen to satisfy Scott’s rule \citep{Scott92}. 

In model~M1, the primary-mass distribution extends to as high as $>100\Msun$ despite the low initial BH mass of $\leq 15\,\Msun$ (Fig.~\ref{fig:spins_8_fidw}~a).
High-mass events with $\gtrsim 50~{\rm M_\odot}$ BHs, including GW190521, primarily arise from high-g mergers as seen in  
Fig.~\ref{fig:masses_gw190521}~a, where 
the distribution is shown for the mergers between high $n$-generation objects with $n\geq4$. 
In model~M1 at $t=10\,\mathrm{Myr}$, 
the expected detection fraction of upper-mass gap mergers with $50\,\Msun\leq m_1 \leq 150\,\Msun$ contributed by $3$, 4, 5, 6, 7, 8, and $n\geq9$th-generation primary BHs 
is 0, 5, 17, 30, 21, 3, and 24 percent, respectively,   
where an $n$g BH signifies a BH 
consisted of $n$ 1g BHs as the merged 
progenitors. 
As the mass of $n$g BHs is at most $\sim n \times 15\,\Msun$ in model~M1, 
primary BHs of GW190521-like mergers $m_1 \sim 85\,\Msun$ come from $n\geq6$-g BHs. 
The fraction of massive binary mergers with $m_\mathrm{bin} \geq130\,\Msun$ ($\sim 90$ percentile lower limit for GW190521) over all events ($f_\mathrm{M130}$) is 0.03 in model~M1. 
Since GW190521 was found among $\sim 60$ triggers, the rate of this type of events can be explained by the AGN channel if it is the dominant channel for all the detected events.
Here, note that the maximum mass of merging binaries ($\sim 185\,\Msun$) is significantly reduced from Papers~I ($\sim 10^4\,\Msun$), which is discussed in the Appendix.

The mergers tend to have high spin around $\sim0.7$ (Fig.~\ref{fig:spins_8_fidw}~b). 
This is because the spin of their members is already high as a result of previous episode of mergers \citep[e.g.][]{Buonanno08}. 
Also, the angles $\theta_1$, $\theta_2$ between the binary orbital angular momentum and the primary and secondary BH spin, respectively, are widely distributed in the entire range of ($0$--$\pi$) (Fig.~\ref{fig:spins_8_fidw}~c) due to several BS interactions, which are assumed to randomize the orientation of the orbital angular momentum.
Yet, strong correlation between $\theta_1$ and $\theta_2$ is still observed owing to spin alignment by gas accretion. 
High-g mergers have a wide range of $\chi_\mathrm{p}$ and $\chi_\mathrm{eff}$ (right panel of Fig.~\ref{fig:masses_gw190521}~a)
due to the spin enhancement during mergers and the 
randomization of the orbital angular momentum directions by BS interactions.

In summary, the results above demonstrate that mergers between high-g BHs in model~M1 is consistent with the observed properties ($\chi_\mathrm{eff}$, $\chi_\mathrm{p}$, $m_1$, $m_2$, $a_1$, $a_2$, $\theta_1$, $\theta_2$, rate) of GW190521. 

Fig.~\ref{fig:masses_gw190521}b shows the parameter distributions of the 2g-2g BH mergers in model~M2, in which the maximum initial BH mass is elevated to $45\,\Msun$, typical for metal-poor progenitors. 
This high initial mass allows 2g-2g BH mergers
to reach the mass range of GW190521. 
The fraction of mergers with $m_\mathrm{bin} \geq130\,\Msun$ becomes as high as 0.11 in this model. 
Since some stars in NSCs are indeed metal-poor \citep{Do20}, 
the observed rate of GW190521-like events is in concordance also with model~M2 
as long as the AGN channel comprises at least $\sim 15\%$ 
of all the GW events detected to date. 
This number is calculated from the observed massive merger fraction ($\sim 1/60$) divided by the fraction in our model ($0.11$). 

Fig.~\ref{fig:masses_gw190521}c shows the results for the 1g-1g mergers in model~M3, in which the accretion rate can be as high as 20 times the Eddington accretion rate. 
Now, owing to rapid growth by accretion, even some 1g-1g mergers reach the mass range of GW190521. 
The gas accretion also significantly enhances the BH spin, and $a_1$, $a_2$, and $\chi_\mathrm{p}$ (or $|\chi_\mathrm{eff}|$).
Such rapid accretion scenario may be favored by the presence of events with high $|\chi_\mathrm{eff}|$ values such as $\chi_\mathrm{eff}=0.81^{+0.15}_{-0.21}$ of GW151216 \citep{Zackay19}. 
Precise estimate for the BH spins in future will further constrain its validity.

In conclusion, the binary property, including the mass, spin and occurence rate, of GW190521 
is consistent with the events via the AGN channel. 
In particular, its origin can be well explained either by high($\geq 4$)-g merger of massive BHs, by 2g BH mergers which have massive low-metallicity progenitor stars, or by mergers of BHs that have rapidly grown by super-Eddington accretion. 

\subsection{lower mass-gap, low mass-ratio, and massive NS mergers}
\label{sec:gw190412}

Next, we examine whether our model can reproduce lower mass-gap, low mass-ratio, and massive NS merger events, like GW190814, GW190412, and GW190425, respectively. 
Fig.~\ref{fig:masses_gw190814} shows the mass distribution of the merging objects for models~M1--M3. 
The mass ranges shown in each panel is chosen so that the distribution around the corresponding observed events is clearly seen. 
We do not show the result of model~M2 for GW190412 since the initial BH mass falls in the range $15$--$45\,\Msun$ by construction and thus 
is in contradiction with the secondary mass of the event ($m_2=8.4^{+1.7}_{-1.0}\,\Msun$). 

In models M1--M3, moderate probability ($f_\mathrm{BHLGO}=0.07$--$0.4\%$, Table~\ref{table_results}) of the detection fraction is predicted for mergers between LGOs and BHs. 
Note that $f_\mathrm{BHLGO}$ can be as high as $\gtrsim 1\%$ depending on the adopted parameters  (see Appendix). 
This is marginally consistent with the rate of events like GW190814, which was found from $\sim 60$ triggers in LIGO/Virgo observations, provided the AGN channel is the dominant channel for all the detected GW events.
Secondary objects of around $2.6\,\Msun$ can be produced by mergers of NSs (panels~a and b), while in model~M3, 1g objects can be as massive as this thanks to super-Eddington accretion.
With low spin magnitude ($\lesssim 0.07$, \citealt{LIGO20_GW190814}), the primary BH of GW190814 is probably 1g and has hardly grown by accretion. 
This fact favors a merger between a massive-1g BH and a LGO as the origin of GW190814 as in model~M2.  

Mergers like GW190412, having the low mass-ratio ($q=0.28^{+0.13}_{-0.07}$) and the high primary-BH spin ($a_1=0.43^{+0.16}_{-0.26}$), 
can be explained by $2$g-$1$g mergers in model~M1 (panel~d of Fig.~\ref{fig:masses_gw190814}) and by 1g-1g mergers in model~M3  (panel~e). 
The detection probability of $2$g-$1$g mergers is $\sim 16\%$ in model~M1, while 
that for low mass-ratio mergers with $q<0.3$ is $\sim 8\%$ in model~M3. 
Also, non-zero spin for the primary BH in GW190412 may be consistent with being $2$g (or high-g) as in model~M1, or experienced significant accretion as in model~M3.

Overall, the properties of GW190814/GW190412 can be explained by the AGN channel either by mergers among $2$g-$1$g objects or mergers of BHs grown by super-Eddington accretion.
The possible 1g-$20\,\Msun$ BH of GW190814 prefers moderately metal-poor progenitors. 
GW190412 allows for solar-metallicity progenitors because high mass of the primary BH ($\sim 30\,\Msun$) suggests it to be a 2 or higher-g object or growth by significant accretion. 
As both high- and low-metallicity stars are present in NSCs \citep{Do20}, AGN disks can provide candidate sites for both events. 

Finally, 
the massive NS merger event GW190425 is well reproduced by growth of the progenitors by gas accretion in model M1 at $t=100\,\mathrm{Myr}$ 
(panel~f of Fig.~\ref{fig:masses_gw190814}).
This result is consistent with the finding in \citet{Yang20_GW190814} that gas accretion expected in the AGN channel can produce GW190425-like massive NS mergers. 
In model M1, the merging NS mass extends to $\sim1.6\,\Msun$ at $t=10\,\mathrm{Myr}$, slightly lower than the CO masses of GW190425.
Since the mass accreted in a fixed time interval is roughly proportional to the gas accretion rate, which is here set to the Eddington rate with $10\%$ radiative efficiency, the mass distribution of NS mergers can constrain the typical accretion rate onto the progenitor and the AGN disk lifetime once a large number of NS mergers in AGNs are observed.  
However, 
the detection fraction of NS-NS mergers over all mergers is $f_\mathrm{NSNS}\sim 5\times 10^{-5}$--$3\times 10^{-3}$ (Table~\ref{table_results}, see also Appendix), which is much lower than the actual observed fraction ($\sim 1/30$). 
Therefore, it might be unlikely that GW190425 is originated from AGNs. 

We have also calculated several models with different sets of the parameters related to the properties of the AGN disk, BH population, etc. and found that the fraction of mass-gap mergers is only moderately affected by the difference in their choice (see Appendix).

\subsection{Influence of assumptions}

Here, we discuss the main assumptions in our model and their possible influence on our main conclusions. 
First, we adopted the 
steady-state 
AGN disk model proposed by \citet{Thompson05} and the initial three-dimensional number density and mass distribution for COs 
as inferred for the  Galactic Center \citep[e.g.][]{Feldmeier14,Schodel14,Do20}.
Since there are uncertainties in the properties of the AGN disks and the CO distributions, 
and they are also expected to vary from one galaxy to another, 
we investigated the dependence of the results on the gas accretion rate from outer radius of the simulation 
(models~M11--M13), the size of the AGN disks (models~M4 and M10), 
the mass of the central SMBH (model~M14), 
the initial mass function of COs (model~M15), the stellar mass enclosed within $3\,\mathrm{pc}$ 
(model~M16), the initial density profile of COs (model~M17), and 
the initial velocity dispersion of COs (models~M5, M18, M19) 
in the Appendix. 

Although the occurrence rates of specific events like mergers moderately depend on 
these parameters
(Table~\ref{table:parameter_model} in the Appendix), 
the range of variations does not significantly affect 
our conclusions that frequent hierarchical mergers in the AGN disks are consistent with 
properties of the observed merger events.
On the other hand, we have not modelled the time evolution of the structure of the AGN disk, 
the formation and evolution of compact objects (and their binaries) other than BHs and NSs, and the possible presence of massive perturbers \citep{Deme20}, which may affect various properties of mergers.

Second, we assumed that COs embedded in the AGN disk migrate toward the SMBH following the formulae for Type I/II migration. However, it is not clear whether radial migration operates 
as prescribed by these planetary formulae,
due to the complexity of the effects of $N$-body migrators \citep{Broz18}, feedback from
BHs \citep{delValle18,Regan19}, and inhomogeneities in the turbulent accretion disk \citep{Laughlin2004,Baruteau10}.
To investigate the impact of the migration on the results, we performed model~M8, in which the radial migration 
is assumed to be inefficient. 
In this model, the number and the maximum mass of mergers are reduced as the rate of binary-single interactions, which facilitate mergers, is strongly enhanced by the migration to the inner regions of the AGN disk, where gaps are opened, and the space density of BHs is high (see Paper~I). 

Third, hierarchical triples composed of three stellar-mass compact objects (e.g. \citealt{Antonini17}; see also Section~5.7.2 of Paper~I), neglected in this study, may influence the spin distribution for some fraction of mergers and the merging mass distribution.

Fourth, we employed a semi-analytical prescription for binary-single interactions. However, this prescription may not be a good approximation for low mass-ratio interactions. 
Also, we assumed that the most massive pair remains in a binary after hard binary-single interactions, and softer binaries are always disrupted after hard binary-binary interactions for simplicity. 
By calculating the interactions using $N$-body simulations, the merging mass distribution may be significantly modified.

As impacts of several of these assumptions on results are uncertain, these should be taken into account in future works.

\section{Conclusions}
In this paper, 
we have followed the evolution of compact objects in AGN disks by way of one-dimensional $N$-body simulations, combined with semi-analytical prescriptions incorporating gaseous and stellar interactions, by extending our previous model to include not only black holes but also neutron stars. 
We have then compared the expected properties of compact-object mergers in AGN disks with those of the observed events GW190521, GW190814, GW190412, and GW190425. 

Our main findings are summarized as follows:
\begin{enumerate}

\item 
The AGN channel can produce high BH-mass mergers like GW190521 in multiple ways: (i) mergers among high-g BHs, (ii) mergers of 2g-2g BHs with metal-poor progenitors, or (iii) mergers of BHs which have acquired mass via moderately super-Eddington accretion. 
Events with roughly equal-mass ratio are expected in this channel, unlike those occurring in the migration traps of the AGN disks.

\item 
Mergers in the lower mass-gap or with unequal masses like
GW190814 and GW190412 can be explained in the AGN channel by 2g-1g mergers or 1g-1g mergers with significant gas accretion, with moderate occurrence probability. 

\item 
The Massive neutron-star merger GW190425 can be explained by the AGN channel if the progenitors had acquired significant mass via accretion, although the predicted rate is much lower than the observed rate.

\end{enumerate}

\acknowledgments

The authors thank Will Farr for useful suggestions. This work is financially supported by the Grants-in-Aid for Basic Research by the Ministry of Education, Science and Culture of Japan (HT:17H01102, 17H06360, KO:17H02869, 17H01102, 17H06360). 
ZH acknowledges support from NASA grant NNX15AB19G and NSF grants 1715661 and 2006176.
This work received founding from the European Research Council (ERC) under the European Union's Horizon 2020 Programme for Research and Innovation ERC-2014-STG under grant agreement No. 638435 (GalNUC) (to BK). J.S. is supported by the
European Unions Horizon 2020 research and innovation programme under the Marie Sklodowska-Curie grant agreement No. 844629. IB acknowledges support from the Alfred P. Sloan Foundation and from the University of Florida. 
Simulations and analyses were carried out on Cray XC50 and computers 
at the Center for Computational Astrophysics, National Astronomical Observatory of Japan.

\bibliographystyle{yahapj.bst}
\bibliography{agn_bhm}

\appendix

\begin{table*}
\begin{center}
	\caption{
	Same as Table~\ref{table_results}, but results for all models. 
     ``No gas hardening'' and ``No gas migration'' respectively represent the models in which the gaseous torques are neglected with respect to the evolution of the binary semi-major axis and the radial position of the binary center of mass within the AGN disk. 
   ``BS in 2D'' and ``BS in 2$\rightarrow$3D'' represents the models 
    in which BS interactions occur in 2D and 2$\rightarrow$3D space, respectively (Paper~III). 
        }
\label{table_results_app}
\begin{tabular}{c|c||c|c|c|c|c|c|c|c}
\hline
\multicolumn{2}{c}{input} \vline& \multicolumn{8}{c}{output}\\\hline
Model&Parameter
&$f_\mathrm{BHLGO}$
&$f_\mathrm{LGOLGO}$
&$f_\mathrm{LGONS}$
&$f_\mathrm{BHBH}$
&$f_\mathrm{BHNS}$
&$f_\mathrm{NSNS}$
&$f_\mathrm{M130}$
&$N_\mathrm{mer}$
\\\hline

M1&Fiducial&
$4\times 10^{-3}$&$3\times10^{-4}$&$3\times 10^{-4}$
&0.989&$6\times 10^{-3}$&$3\times 10^{-4}$
&$0.03$
&$1.6\times 10^3$
\\\hline

M2&$m_\mathrm{1gBH}~\times =3$&
$7\times 10^{-4}$&$5\times 10^{-5}$&$7\times 10^{-5}$
&0.998&$1\times 10^{-3}$&$5\times 10^{-5}$
&0.11
&$2.0\times 10^3$
\\\hline

M3&$\Gamma_\mathrm{Edd,cir} =20$&
$5\times 10^{-3}$&$5\times 10^{-5}$&$9\times 10^{-5}$
&0.994&$8\times 10^{-4}$&$5\times 10^{-5}$
&0.35
&$1.9\times 10^3$
\\\hline

M4&$r_\mathrm{out,BH}=0.03\,\mathrm{pc}$&
$0.10$&$2\times 10^{-3}$&$3\times 10^{-3}$
&0.82&$0.07$&$3\times 10^{-3}$
&0
&$3.1\times 10^2$
\\\hline

M5&${\beta}_\mathrm{v,BH}={\beta}_\mathrm{v,NS}=1$&
$0.03$&$2\times 10^{-3}$&$2\times 10^{-3}$
&0.94&$0.03$&$2\times 10^{-3}$
&0
&$5.9\times 10^2$
\\\hline

M6&$f_\mathrm{LGO}=0.1$&
$0.01$&$1\times 10^{-3}$&$7\times 10^{-4}$
&0.984&$4\times 10^{-3}$&$4\times 10^{-4}$
&0.03
&$1.8\times 10^3$
\\\hline

M7&$f_\mathrm{LGO}=0.01$&
$6\times 10^{-3}$&$3\times 10^{-4}$&$5\times 10^{-4}$
&0.989&$5\times 10^{-3}$&$4\times 10^{-4}$
&0.01
&$1.6\times 10^3$
\\\hline

M8&No gas mig.&
$0.02$&$1\times 10^{-3}$&$2\times 10^{-3}$
&0.95&$0.03$&$2\times 10^{-3}$
&0
&$7.1\times 10^2$
\\\hline

M9&No gas hard.&
$2\times 10^{-3}$&$7\times 10^{-4}$&$2\times 10^{-3}$
&0.98&$0.01$&$1\times 10^{-3}$
&0.11
&$1.5\times 10^3$
\\\hline

M10&$r_\mathrm{out,BH}=0.3\,\mathrm{pc}$&
$8\times 10^{-3}$&$7\times 10^{-4}$&$1\times 10^{-3}$
&0.98&$0.01$&$8\times 10^{-4}$
&0.09
&$1.0\times 10^3$
\\\hline

M11&${\dot M}_\mathrm{out}={\dot M}_\mathrm{Edd}$&
$6\times 10^{-3}$&$2\times 10^{-4}$&$5\times 10^{-4}$
&0.988&$5\times 10^{-3}$&$4\times 10^{-4}$
&0.06
&$1.7\times 10^3$
\\\hline

M12&${\dot M}_\mathrm{out}=0.01{\dot M}_\mathrm{Edd}$&
$2\times 10^{-3}$&$5\times 10^{-4}$&$6\times 10^{-4}$
&0.992&$4\times 10^{-3}$&$7\times 10^{-4}$
&0.01
&$8.9\times 10^2$
\\\hline

M13&${\dot M}_\mathrm{out}=0.001{\dot M}_\mathrm{Edd}$&
$3\times 10^{-3}$&$0$&$4\times 10^{-4}$
&0.98&$0.02$&$1\times 10^{-3}$
&0
&$3.1\times 10^2$
\\\hline

M14&$M_\mathrm{SMBH}=4\times 10^7\Msun$&
$8\times 10^{-3}$&$2\times 10^{-4}$&$4\times 10^{-4}$
&0.98&$7\times 10^{-3}$&$3\times 10^{-4}$
&0.07
&$1.6\times 10^3$
\\\hline

M15&$\delta_\mathrm{IMF}=1.7$&
$1\times 10^{-3}$&$1\times 10^{-4}$&$1\times 10^{-4}$
&0.997&$1\times 10^{-3}$&$1\times 10^{-4}$
&0.09
&$8.3\times 10^3$
\\\hline

M16&$M_\mathrm{star,3pc}=3\times 10^6\Msun$&
$0.01$&$2\times 10^{-4}$&$6\times10^{-4}$
&0.98&$0.01$&$5\times 10^{-4}$
&0
&$5.3\times 10^2$
\\\hline

M17&$\gamma_\mathrm{\rho,BH}=1.5$&
$1\times 10^{-3}$&$0$&$7\times10^{-5}$
&0.994&$4\times 10^{-3}$&$2\times 10^{-4}$
&0
&$9.8\times 10^2$
\\\hline

M18&${\beta}_\mathrm{v,BH}=0.4$&
$9\times 10^{-3}$&$1\times 10^{-3}$&$2\times 10^{-3}$
&0.97&$0.01$&$1\times 10^{-3}$
&0
&$1.2\times 10^3$
\\\hline

M19&${\beta}_\mathrm{v,NS}=0.2$&
$6\times 10^{-3}$&$7\times 10^{-4}$&$8\times 10^{-4}$
&0.987&$6\times 10^{-3}$&$7\times 10^{-4}$
&0.06
&$2.2\times 10^3$
\\\hline

M20&BS in 2D&
$3\times 10^{-3}$&$1\times 10^{-4}$&$6\times 10^{-4}$
&0.987&$9\times 10^{-3}$&$3\times 10^{-4}$
&0.07
&$4.6\times 10^3$
\\\hline

M21&BS in $2\rightarrow3$D&
$3\times 10^{-3}$&$2\times 10^{-4}$&$4\times 10^{-4}$
&0.991&$6\times 10^{-3}$&$3\times 10^{-4}$
&0.05
&$3.4\times 10^3$
\\\hline

M22&$m_\mathrm{1gBH} =10\,\Msun$&
$3\times 10^{-3}$&$3\times 10^{-4}$&$3\times 10^{-4}$
&0.991&$5\times 10^{-3}$&$4\times 10^{-4}$
&$0.01$
&$1.6\times 10^3$
\\\hline

M1&$t=1\,\mathrm{Myr}$&
$4\times 10^{-3}$&$3\times10^{-4}$&$9\times 10^{-4}$
&0.986&$8\times 10^{-3}$&$8\times 10^{-4}$
&$0$
&$2.6\times 10^2$
\\\hline

M1&$t=100\,\mathrm{Myr}$&
$0.02$&$9\times 10^{-4}$&$1\times 10^{-3}$
&0.96&$0.02$&$8\times 10^{-4}$
&$8\times 10^{-3}$
&$7.3\times 10^3$
\\\hline

M12&${\dot M}_\mathrm{out}=0.01{\dot M}_\mathrm{Edd}, t=100\,\mathrm{Myr}$&
$7\times 10^{-3}$&$5\times 10^{-4}$&$1\times 10^{-3}$
&0.98&$7\times 10^{-3}$&$7\times 10^{-4}$
&0.05
&$4.6\times 10^3$
\\\hline

M12&${\dot M}_\mathrm{out}=0.01{\dot M}_\mathrm{Edd}, t=1\,\mathrm{Gyr}$&
$4\times 10^{-3}$&$2\times 10^{-3}$&$6\times 10^{-3}$
&0.92&$0.03$&$3\times 10^{-3}$
&0.07
&$1.7\times 10^4$
\\\hline

M13&${\dot M}_\mathrm{out}=0.001{\dot M}_\mathrm{Edd}, t=100\,\mathrm{Myr}$&
$3\times 10^{-3}$&$4\times 10^{-4}$&$5\times 10^{-4}$
&0.993&$3\times 10^{-3}$&$4\times 10^{-4}$
&$4\times 10^{-3}$
&$2.2\times 10^3$
\\\hline

\end{tabular}
\end{center}
\end{table*}

\section{Alternative models}
\label{sec:other_models}

In this section, we present the parameter dependence of the results using a total of 22 different models, listed in Table~\ref{table_results_app}. We also discuss merging mass distributions in the AGN channel. 

\section{Mergers in upper-mass gap}

We find that the fraction of mergers of massive binaries of $\geq130\,\Msun$ ($f_\mathrm{M130}$) is zero when the BH number in inner regions ($r\lesssim 0.01\,\mathrm{pc}$) of the AGN disk is low due to the small size of the AGN disk (model~M4), the high velocity dispersion of BHs (models~M5, M16), no radial migration (model~M7), low mass for NSCs (model~M14), shallow density profile of BHs (model~M15), or short duration of an AGN phase ($t=1\,\mathrm{Myr}$, model~M1). 
On the other hand, $f_\mathrm{M130}$ is enhanced if the initial masses of 1g BHs are massive (model~M2), super-Eddington accretion is allowed (model~M3), large number of BHs initially exists (model~M13), or BHs can efficiently migrate to inner regions of $r\lesssim10^{-3}\,\mathrm{pc}$. 
In models~M8--M11, the fraction of mergers in the inner regions ($r\lesssim10^{-3}\,\mathrm{pc}$) is higher than that in model~M1. 
In the inner regions, massive BHs can merge with high binary eccentricity due to GW-capture binary formation 
(Paper~III, see also \citealt{Gondan20_GW190521})
which may be consistent with possible high eccentricity in GW190521 implied by \citet{Gayathri20_GW190521}.

Here, the maximum masses of merging binaries in the fiducial model ($\sim 185\,\Msun$) are significantly reduced from those in Papers~I ($\sim 10^4\,\Msun$). This is due to the exchange interactions and disruption of softer binaries during binary-binary interactions, which are not included previously.  With the exchanges included, both the number and maximum mass of mergers become smaller, because massive objects are retained in binaries after the exchanges and the more energy must be extracted for the more massive binaries to merge, during which many lighter COs receive strong kicks, reducing the number of COs in AGN disks. 

Also, in our fiducial model, mergers between two high-generation BHs are common 
and the majority of them have high mass-ratio close to unity, as is the case for GW190521. Such high-g-high-g mergers do not occur in the model by \cite{Yang19a}, who assumed that mergers occurs in the migration traps shortly after being captured in the disk. In that case, runaway growth of a single object by $n$g-$1$g mergers took place instead. In AGN disks outside the migration traps, high-g-high-g mergers can be more frequent because of the following reasons: (i) it takes longer for binary hardening and merger, (ii) objects kicked out from the disk by BS interactions interact less with others, and (iii) different objects orbit and merge at different distances from the SMBH due to variations in initial positions, gap-opening regions, and migration speeds. Combinations of these enable objects to avoid merging into one massive object in contrary to mergers in the migration traps.

In the fiducial model, 
at $t=1$, $10$, and $100~\mathrm{Myr}$, respectively, 
the maximum masses of merging binaries are $130$, $185$, and $185~\Msun$, 
and the numbers of mergers are $260$, $1,600$, and $7,300$, if the AGN disk exists for these durations. 
Thus, the merger rate decreases with time, as the rate of new BH captures in the disk by gas dynamical friction decreases.
This is because of the strong dependence of the capture timescale on 
the initial CO velocity relative to the local AGN disk gas motion ($\propto v_\mathrm{ini}^4$, Eq.~26 in paper~I). COs having low $v_\mathrm{ini}$ are quickly captured, while those with high $v_\mathrm{ini}$ are typically not captured by the AGN disk within $100\,\mathrm{Myr}$. 
Also, the maximum masses are not sensitive to disk lifetimes beyond $t\geq10\,\mathrm{Myr}$. 
This is because mergers become infrequent and 
massive BHs often migrate to the inner boundary. 
Here, the number of BHs that migrate to the inner boundary of $r=10^{-4}\,\mathrm{pc}$ is $\sim90$ at $t=100\,\mathrm{Myr}$, while it is $\sim10$ at $t=10\,\mathrm{Myr}$. 
Note that objects which reach $r\leq10^{-4}\,\mathrm{pc}$ are not followed in our model, and are instead removed from the simulation. In these regions, COs may merge with each other if they accumulate in migration traps \citep{Bellovary16} or they may accrete onto the SMBH and observed by LISA as the migration timescale due to GW emission is 
comparable to the AGN lifetime ($\sim 30\,\mathrm{Myr} (r/10^{-4}\,\mathrm{pc})^{4}(M_\mathrm{SMBH}/4\times 10^6\,\Msun)^{-2}(m_\mathrm{CO}/100\,\Msun)^{-1}$, where $M_\mathrm{SMBH}$ and $m_\mathrm{CO}$ are the SMBH and the CO masses). 
Note that as massive BHs are rare, the evolution of maximum masses likely suffers some variations. For example, with an independent realization of the initial condition, the maximum masses of merging binaries are $70$, $138$, and $186~\Msun$, at $t=1$, $10$, and $100~\mathrm{Myr}$, respectively. 
In all cases, the growth of the maximum mass is slowed down in later phases.

In the fiducial model, $f_\mathrm{M130}$ decreases in $t\geq10\, \mathrm{Myr}$ due to inefficient capture of BHs in later phases and migration of massive BHs to the inner boundary of the AGN disk as discussed above. 
On the other hand, $f_\mathrm{M130}$ increases in $t\geq10\, \mathrm{Myr}$ in model~M12 and M13, in which the gas accretion rate at the outer boundary is low (${\dot M}_\mathrm{out}=0.01$ and $0.001\,{\dot M}_\mathrm{Edd}$). 
In addition, in model~M12, 
the maximum mass of merging binaries increases with time as 
those are $46$, $142$, $228$, and $283~\Msun$ at $t=1$, $10$, $100$, and $1000~\mathrm{Myr}$, respectively. 
Thus, the maximum merging mass and $f_\mathrm{M130}$ evolve more in later phases for low ${\dot M}_\mathrm{Edd}$ models 
compared to that in the fiducial model. 
These are due to 
the difference of the gas density of the AGN disk. 
Inefficient gaseous processes (capture to the AGN disk and migration) due to low gas density facilitate growth of BHs in later phases. 
Such low ${\dot M}_\mathrm{out}$ cases may be more important for understanding properties of mergers in the AGN channel 
as the number of mergers from AGNs with low Eddington rates may dominate over those with high Eddington rates depending on the Eddington rate distribution of AGNs \citep{Hopkins_2005ApJ...630..705H,Greene07,Shankar13}, which is uncertain. 
If the merger rate is dominated by AGNs with low Eddington rates, 
the rate is significantly influenced by the transition radius from a standard disk to advection-dominated accretion flow, which possibly comes to $\gtrsim10^{-3}$--$10^{-2}\,\mathrm{pc}$ and affect merger processes for ${\dot M}_\mathrm{out}\lesssim 10^{-3}\,{\dot M}_\mathrm{Edd}$ (Eq.~3 of \citealt{Czerny19}), although the typical value of the radius is uncertain \citep[e.g.][]{Czerny19}.

\section{Mergers in lower mass gap}

The fraction of BH-LGO mergers is enhanced by several effects. 
In model~M4, we assume that COs initially exist within $r_\mathrm{out,BH}=0.03\,\mathrm{pc}$. 
This resembles the situation in which the size of the AGN disk is $0.03\,\mathrm{pc}$. 
In this model, $f_\mathrm{BHLGO}$ is enhanced to $0.10$. 
In the inner regions of the AGN disk, 
the capture timescale of COs by the AGN disk is shorter. As a large fraction of NSs and BHs are rapidly captured, the ratio of the number of NSs over BHs in the AGN disk is high and accordingly the rates of NS-NS and BH-LGO mergers are also relatively high.

In model~M5, in which the initial velocity dispersion of BHs and NSs is virial ($\beta_\mathrm{v,BH}=\beta_\mathrm{v,NS}=1$), 
$f_\mathrm{BHLGO}$ is enhanced to $0.03$. 
As the initial velocity dispersion of BHs and NSs increases, the numbers of BHs and NSs in the AGN disks decrease. Then, binary NSs have a high probability to merge without interacting with BHs and experiencing exchange of binary components. 
The high velocity dispersion of BHs might be realized if the AGN phase begins before dynamics of NSCs is regulated by vector resonant relaxation e.g. due to randomization of orbits of COs by major merger of galaxies, or if NSCs have negligible rotation on average.

In models~M6 and M7, we assume that NSs have already merged prior to the AGN phase. Such situation may be realized if the high rate of NS-NS mergers ($\sim 1000\,\mathrm{Gpc^{-3}yr^{-1}}$) inferred from GW observations are driven by isolated binary evolution \citep[but see][]{Belczynski18} and such mergers are not prevented by soft-BS interactions in NSCs, or NS mergers are facilitated by the Kozai-Lidov mechanism \citep[e.g.][]{Fragione19_NSBH_GN_I}. 
To investigate the effects of this possibility, we assume that a fraction of NSs ($f_\mathrm{LGO}$) already merged with NSs prior to the start of the AGN episode, and LGOs may have initial masses with $2.6\,\Msun$. 
We investigate a model with $f_\mathrm{LGO}=0.1$ (M6), which would present an upper limit as the rate of core-collapse supernovae is $\sim 10^5\,\mathrm{Gpc^{-3}yr^{-1}}$ \citep{Madau14} and that of NS-NS mergers is $\sim 100$--$4000\,\mathrm{Gpc^{-3}yr^{-1}}$ \citep{TheLIGO18}. 
$f_\mathrm{BHLGO}$ is moderately enhanced to $0.01$ for $f_\mathrm{LGO}=0.1$ from its value of $4\times 10^{-3}$ for $f_\mathrm{LGO}=0$. 
Thus, isolated binary evolution may contribute to the fraction of mergers in the lower mass gap ($f_\mathrm{BHLGO}$) in the AGN channel.

In model~M8, in which COs do not migrate toward the SMBH, $f_\mathrm{BHLGO}$ is higher as the probability that the NS binaries interact with BHs and the binary components are exchanged is reduced. 
The efficiency of migration of COs in AGN disks is uncertain, while it affects various properties of mergers in AGN disks. Thus, this process is worth investigating in more detail in future work to better understand the implications for mergers in AGN disks.

\begin{figure*}\begin{center}
\includegraphics[width=160mm]{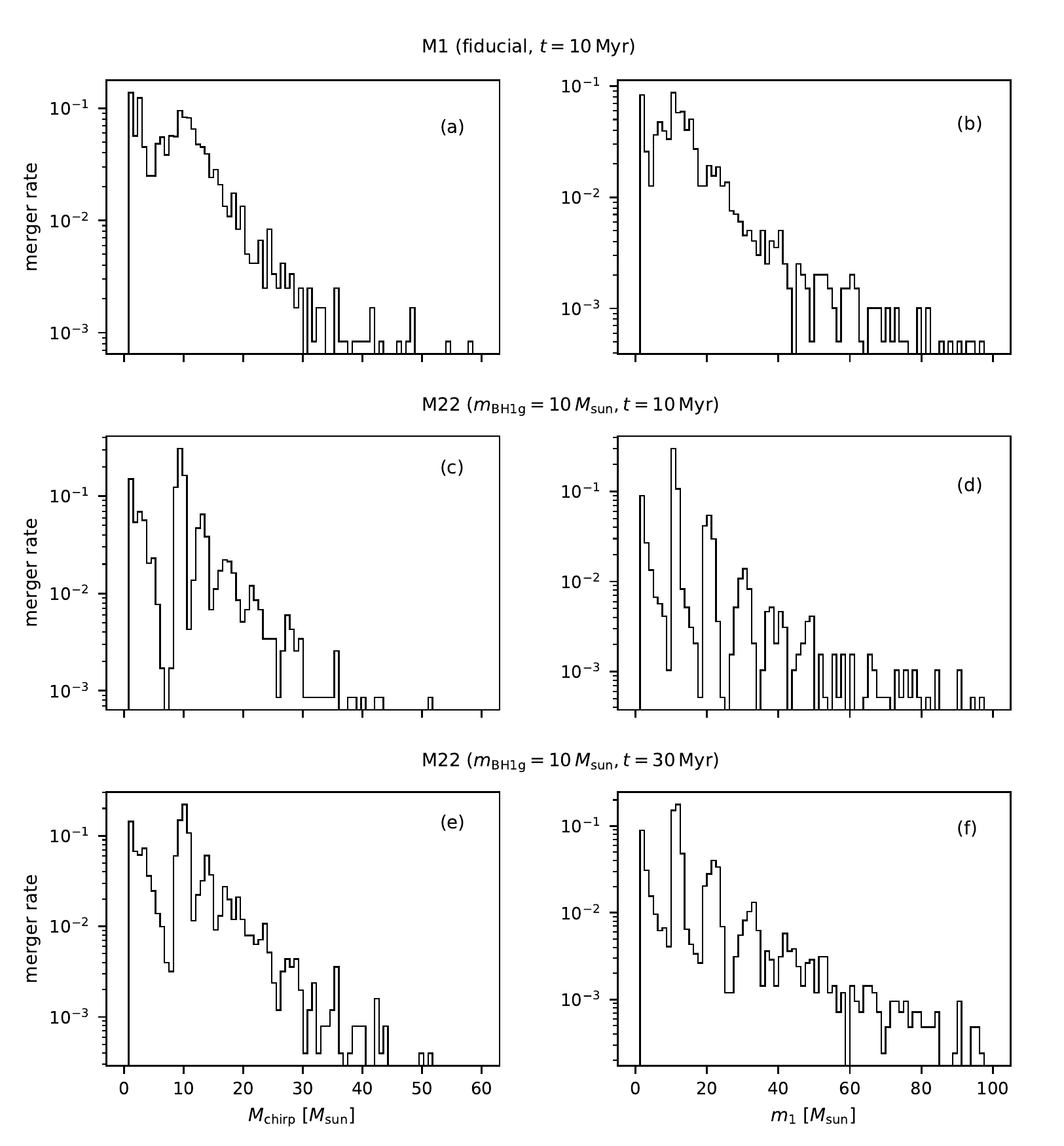}
\caption{
The distribution of the chirp (left) and primary (right) masses 
for model~M1 at $t=10\,\mathrm{Myr}$ (panels~a and b) and 
model~M22 at $t=10\,\mathrm{Myr}$ (panels~c and d) 
and $t=30\,\mathrm{Myr}$ (panels~e and f). 
}\label{fig:m_dist}
\end{center}\end{figure*}

\section{Merging mass distribution}

Here, we discuss the merging mass distribution. 
As a large number of events are reported in LIGO/Virgo O3a \citep{LIGO20_O3_Catalog}, we can assess the consistency 
between predicted and observed distributions more precisely. 
Recently, \citet{LIGO20_O3_Properties} and \citet{Tiwari20} analyzed the LIGO/Virgo's O1-O3a data and reconstructed merging mass distributions. 
\citet{LIGO20_O3_Properties} reconstructed the primary mass distribution by a superposition of a (broken) power law and $0$--$2$ Gaussian distributions, while \citet{Tiwari20} reconstructed the chirp mass distribution by a superposition of twelve Gaussian distributions, which can flexibly recover features of the mass distribution. 
\citet{LIGO20_O3_Properties} suggested that 
the $m_1$ distribution is more consistent with a broken power law with a break at $m_1 \sim 40\,\Msun$, or a power law with a Gaussian feature peaking at $m_1\sim34\,\Msun$. 
Interestingly, \citet{Tiwari20} identified four peaks in the chirp mass distribution at 8, 14, 26, and $45\,\Msun$, which are roughly separated by a factor of around two, possibly suggesting frequent hierarchical mergers. 
Here, high values for $\chi_\mathrm{p}$ inferred for observed GW events \citep{Gerosa20_xp} may also support this scenario. 
As hierarchical mergers are presumably most frequent in the AGN channel, it would be interesting to check whether such features can be reproduced in our model. 

In model~M1, 
peak-like structures are less clear in the chirp mass distribution (Fig.~\ref{fig:m_dist}~a) despite of frequent hierarchical mergers. 
On the other hand, as the merging mass distribution is significantly affected by the adopted initial mass function of BHs, we performed model~M22 in which the initial masses of all 1g BHs are set to $10\,\Msun$. Such a spiky initial mass function may be allowed depending on prescriptions of stellar mass loss \citep{Belczynski10}. 
In this model, peaks in the $m_\mathrm{chirp}$ distribution become much clearer (panels~c and e), although the number of peaks is larger than that implied by \citet{Tiwari20}, due to the contribution of mergers among various generation BHs like 1g-2g or 2g-3g. 
This model predicts that further peaks might be detectable by observing a larger number of events. On the other hand, the contribution of mergers among 1g-2g or 2g-3g BHs is supposedly influenced by frequency of BS interactions, which depends on several parameters and processes, such as $\delta_\mathrm{IMF}$, $\beta_\mathrm{v,BH}$, and the efficiency of migration due to gaseous torque. 
Also, some peaks become smooth and close to a power-law like shape if we adopt different initial mass functions for BHs as in model~M1 (panel~a) or as time passes (panels~c and e). Observational errors on the chirp mass further make the merging mass distribution somewhat smoother. 
As properties of peaks are influenced by several parameters, uncertainties, and observational effects, the features inferred by \citet{Tiwari20} might be reproduced. 

Meanwhile, the $m_1$ distributions (Fig.~\ref{fig:m_dist}~b, d, and e) look roughly similar to those reconstructed by \citet{LIGO20_O3_Properties}. Note that the errors on $m_1$ are typically much larger than that on the chirp mass, which presumably reduces peaks in the observed $m_1$ distribution (Fig.~\ref{fig:m_dist}~d and f). 
Hence, there is a possibility that the AGN channel can explain the observed features of merging masses. 
To constrain the contribution from the AGN channel, quantitative comparisons would be required, which will be conducted in a following work.

\begin{table*}
\begin{center}
\caption{Fiducial values of our model parameters. 
}
\label{table:parameter_model}
\hspace{-5mm}
\begin{tabular}{c|c}
\hline 
Parameter & Fiducial value \\
\hline\hline
Spacial directions in which BS interactions occur&  isotropic in 3D\\\hline
Number of temporary binary COs formed during a BS interaction&  $N_\mathrm{int}=20$\\\hline
Initial BH spin magnitude & $|{\bm a}|=0$\\\hline
Angular momentum directions of circum-CO disks & $\hat{{\bm J}}_\mathrm{CCOD}=\hat{{\bm J}}_\mathrm{AGN}$ for single COs,\\&
$\hat{{\bm J}}_\mathrm{CCOD}=\hat{{\bm J}}_\mathrm{bin}$ for COs in binaries\\\hline
Ratio of viscosity responsible for warp propagation\\over that for transferring angular momentum 
& $\nu_2/\nu_1=10$ \\\hline
Alignment efficiency of the binary orbital angular momentum
due to gas capture 
& $f_\mathrm{rot}=1$
\\
(Eq.~14 in Paper~II)&\\
\hline 
Mass of the central SMBH & $M_\mathrm{SMBH}=4\times 10^6\,\Msun$ \\\hline
Gas accretion rate at the outer radius of the simulation ($5\,\mathrm{pc}$)
& ${\dot M}_\mathrm{out}=0.1\,{\dot M}_\mathrm{Edd}$ with $\eta=0.1$\\\hline
Fraction of pre-existing binaries & $f_\mathrm{pre}=0.15$ \\\hline
Power-law exponent for the initial density profile for BHs and NSs & $\gamma_\mathrm{\rho,BH}=0$, $\gamma_\mathrm{\rho,NS}=-0.5$ \\\hline
Initial velocity anisotropy parameter\\such that $\beta_\mathrm{v,BH}v_\mathrm{kep}(r)$ is the BH velocity dispersion  
& $\beta_\mathrm{v,BH}=0.2$, $\beta_\mathrm{v,NS}=0.4$ \\\hline
Efficiency of angular momentum transport in the $\alpha$-disk & $\alpha_\mathrm{SS}=0.1$ \\\hline
Stellar mass within 3 pc &$M_\mathrm{star,3pc}=10^7\,\Msun$\\\hline 
Stellar initial mass function slope & $\delta_\mathrm{IMF}=2.35$\\\hline
Angular momentum transfer parameter in the outer star forming regions 
&$m_\mathrm{AM}=0.15$\\
(Eq.~C8 in \citealt{Thompson05}) &\\
\hline
Accretion rate in Eddington units onto\\stellar-mass COs with a radiative efficiency $\eta=0.1$
&$\Gamma_\mathrm{Edd,cir}=1$\\\hline
Numerical time-step parameter &$\eta_t=0.1$\\\hline
Number of radial cells storing physical quantities &$N_\mathrm{cell}=120$\\\hline
Maximum and minimum $r$ for the initial CO distribution&  $r_\mathrm{in,BH}=10^{-4}$ pc, $r_\mathrm{out,BH}=3$ pc \\\hline
Initial number of BHs and NSs within 3 pc &$N_\mathrm{BH,ini}=2\times 10^4$, $N_\mathrm{NS,ini}=5\times 10^4$ \\\hline
\end{tabular}
\end{center}
\end{table*}

\end{document}